\documentclass[conference]{IEEEtran}
\usepackage[utf8]{inputenc}
\usepackage[nolist]{acronym}
\usepackage{amsmath}
\usepackage[table,xcdraw]{xcolor}
\usepackage{balance}
\usepackage{booktabs,tabularx}
\newcolumntype{L}{>{\centering\arraybackslash}m{3cm}}
\usepackage[pdftex]{graphicx}
\usepackage{enumitem}
\usepackage[font=small,skip=0pt]{caption}
\begin{document}
\title{Over the Sea UAV Based Communication} %UAV communication for search and rescue  over the sea: is it feasible?
\author{\IEEEauthorblockN{Gianluca Fontanesi\IEEEauthorrefmark{1}, Hamed Ahmadi\IEEEauthorrefmark{1}\IEEEauthorrefmark{2}, and Anding Zhu\IEEEauthorrefmark{1}}
\\
\IEEEauthorrefmark{1}School of Electrical and Electronic Engineering, University College Dublin, Ireland\\
\IEEEauthorrefmark{2}School of Computer Science and Electronic Engineering, University of Essex, United Kingdom
}

\maketitle

\begin{abstract}
     \ac{UAV} aided wireless networks have been recently envisioned as a solution to provide a reliable, low latency cellular link for search and rescue operations over the sea. We propose three different network architectures, based on the technology deployed on the UAV: a flying relay, a flying \ac{BS} and a flying \ac{RRH}. We describe the challenges and highlight the benefits of the proposed architectures from the perspective of search and rescue operations over the sea. We compare the performance in term of data rate and latency, analyzing different solutions to provide a \ac{BH}/\ac{FH} link for long coverage over the sea. Results show that a system architecture is not outperforming over the others. A cost function is thus indicated as a tool to find a suboptimal solution.
\end{abstract}


\section{Introduction}
% \ac{UAV} communication is an hot and promising topic in the \ac{5G} of wireless networks. Researchers and universities are putting much effort to include flying platforms of various type and size in the \ac{5G} and to demonstrate their effective utility in several applications and use cases \cite{MOZ_SAAD_BENNIS_EfficientDeploymentUAV} \cite{Prague_Uni_Gesbert_CanUAVsubstituteSCBS}. As a result, 3GPP is planning to support non terrestrial networks, i.e. \acp{UAV}, in the second phase of the new radio standard which is expected to appear in the 3GPP Rel-16 by mid-2019\footnote{http://www.3gpp.org/DynaReport/WI-List.htm}.

%Everyday news highlights the migrant crisis in Europe where Mediterranean crossings are deadlier than ever. Since 2014,  more  than  14,500  people have lost their lives in their attempt to reach  Europe's  shores \cite{IOM1}. 

Effective and efficient offshore search and rescue operations can save many lives. The significant increase in the number of small boats attempting to cross Mediterranean sea and lives lost on these attempts highlighted importance of these operations \cite{CUSUMANO201791}. Information sharing and communication among rescue vessels and rescuees, will significantly improve the efficiency and effectiveness of search and rescue operation over the sea. A reliable and robust communication system is, therefore, a crucial asset for search and rescue operations over the sea. 

%harsh and limited by the lack of proper communication infrastructures.
Existing satellite services are expensive and suffer from high latency. Moreover, small (unregistered) vessels normally do not have satellite mobile phones onboard. Common European standards, specifically designed for use in emergency, e.g. \ac{TETRA}, have a very slow data transfer. % (typically 80 kbit/s to 250 kbit/s). 
In this paper, we are motivated by the need of investigating new technologies to realize an effective communication system in an event of emergency over the sea for enhancing timely rescue and recovery operations. 
%\footnote{https://www.etsi.org/technologies-clusters/technologies/tetra}
In a recent work \cite{Fonta_UAVSea}, we presented two different \ac{UAV}-assisted wireless network architectures as promising solutions to extend the coverage of the terrestrial network and assist the rescue and recovery operations.
\acp{UAV} introduce a high level of mobility and flexibility to the network and can be deployed where and when they are needed. \acp{UAV} can complement existing cellular networks when additional capacity or coverage is needed, e.g. in outdoor music concerts or sport events \cite{Zhang_UAV_opportunitiesChallenges}.
In addition, given the high dynamism, \acp{UAV} are envisioned as the immediate responder to resurrect the communication link in post disaster scenarios (e.g. project ABSOLUTE). % \cite{ABSOLUTE}). 
Beyond the benefits to the network, using \acp{UAV} to provide connectivity for search and rescue operations over the sea is important to prevent serious losses, for example, by saving lives. In fact, high quality communication between rescuers and people in distress enable applications like videos streaming, users location and monitoring/recognition. Even more attractive from a search and rescue point of view,  one or more \acp{UAV} attached to the network as flying \acp{UE} could act as mechanical eyes to monitor wide sea areas. For this reason, we believe not enough attention has been given to \ac{UAV} as communication provider for search and rescue operations over the sea.

We classify \ac{UAV}-assisted networks based on the role of \ac{UAV} into three main groups:
1) \ac{UAV} carries a \ac{BS}, 2) \ac{UAV} is loaded only with the radio front end or \ac{RRH} and, 3) \ac{UAV} acts as a mobile relay.

Deployed as flying \acp{BS} in the sky, \acp{UAV} are able to adjust their locations dynamically and provide on demand services to the ground users according to their real-time locations. % \cite{MOZ_SAAD_BENNIS_EfficientDeploymentUAV}.
This first approach has been widely investigated, especially for emergency and post disaster scenarios.
For instance, the work in \cite{Murphy_Sreenan_RescueMountainUAV} suggests low cost \acp{UAV} for search and rescue operations in the mountain to survey and locate individuals in distress.
Moreover, the flying \acp{BS}  architecture has been proposed for search and rescue operations in deep open pit mines \cite{Ranjan_UAV_Mines} or disaster struck scenarios \cite{Disaster_UANetwrok_IEEEMAgazine}.
% Given the endemic \ac{UAV} limitation in energy and payload capacity, hardware cost and weight limit the deployment of this approach.
% Further, the altitude of the \acp{UAV} is usually making the favorable \ac{LOS} channel available for the ground users without deep fading \cite{OptimalLAPAltitude}. The work in \cite{Murphy_Sreenan_RescueMountainUAV} suggests low cost \acp{UAV} for search and rescue operations in the mountain to survey and locate individuals in distress. The authors in \cite{8255735} used \acp{UAV} in disaster-resilience where they present a disaster struck scenario where they presented the trade-off between the altitude, beamwidth angles and the coverage area of the \acp{UAV}.

Using a \ac{UAV} to relay the cellular networks can reduce the weight on  \ac{UAV}'s payload while extending the ground cellular coverage to remote and dedicated areas where infrastructure is not available or expensive to deploy \cite{Ahmadi_SON_airborne}.  %\cite{WirelessRelay_UAV_Zhan}. 

\acp{UAV} can be integrated into mobile networks acting as flying \ac{RRH} in similar way as in a \ac{C-RAN} network, where in the original terrestrial configuration several fixed \acp{RRH} serve as access points to the network and are connected to a \ac{BBU} that performs the major part of the processing \cite{AHmadiBBU_RRH_distance}.
Flying \acp{RRH} could be deployed to increase channel quality, user throughput and the energy efficiency of the system \cite{Prague_Uni_Gesbert_CanUAVsubstituteSCBS},
%The \ac{C-RAN} concept allows to exploit the processing capacity available in the cloud, as well to achieve load balancing and reuse of processing resources \cite{AHmadiBBU_RRH_distance} 
but it introduces challenges that range from the high capacity required on the \ac{FH}, to strict latency and jitter requirements. %M

% However, to be considered a real alternative, this configuration, in our opinion, must be provided of a \ac{FH} technology able to manage the huge load of data that can be imposed on the wireless \ac{FH} link.

The rising number of works considering the aforementioned deployment approaches motivates us to investigate the benefits and challenges of these technologies while deployed on \acp{UAV}.
In this paper, we take a new look at the challenges and appealing features of the approaches presented above, from the perspective of search and rescue operations over the sea. Specifically, the contributions of this paper are: an understanding of challenges and benefits of the three UAV assisted network architectures in a realistic search and rescue scenario over the sea; a performance comparison based on different indicators; a cost function to design the suboptimal approach considering different \ac{BH} technologies. 
% an highlight of the main limitation for each single approach.

% Different from any other work, to the best of our knowledge, we investigate then what and which of three system approaches could suit better this scenario.

% The promising features of the approaches presented here above motivates this paper to investigate an open question, so far not answered: is \ac{UAV}-\ac{BS} configuration suitable for an emergency use case like search and rescue over the sea? And what for a \ac{UAV}-\ac{RRH}? Even if the concept of using \acp{UAV} for emergency operations is not novel, this work is the first to the best of our knowledge that illustrate the benefits of this approach for the specific over the sea scenario.

\section{System architectures description and related challenges}
In most cases, search and rescue operations for migrants at sea usually take place off-shore \cite{CUSUMANO201791}. In this area, typically, the cellular coverage is very low or absent.
This lack motivates to deploy a wireless \ac{UAV}-aided wireless network to provide temporary cellular connectivity for all the entire length of the search and rescue operations. The network is composed by a \ac{UAV} that can act as flying relay, flying \ac{BS} or flying \ac{RRH} to serve the users.
Based on the technology deployed on the \ac{UAV}, three different end-to-end network architectures can be considered:

\begin{enumerate}[label=\Roman*]
    \item Core Network - Ground Base Station - Relay - Users
    \item Core Network - Ground Base Station - Flying Base Station - Users
    \item Core Network - Base Band Unit - Flying Remote Radio Head - Users
\end{enumerate}

\begin{figure}
    \centering
    \includegraphics[width=1\columnwidth]{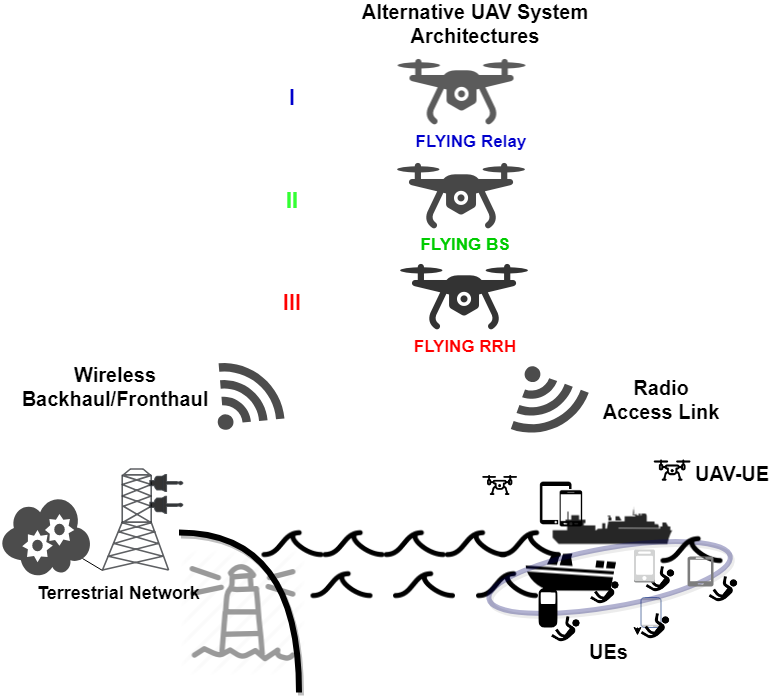}
    \caption{Graphical illustration of the architecture for extending the cellular coverage over the sea using flying BS, RRH and relays}
    \label{fig:SystemMOdel}
\end{figure}

In the first system architecture, we consider the UAV to work as a relay type 0, where under amplify and forward scheme, half of time slots are assigned to transmit the signal to the UAV and the second half to listen the signal from the ground BS.
The solutions proposed with the architectures II and III require, respectively, a wireless \ac{BH} link and wireless \ac{FH} link to the core network.
%The \ac{UAV} has limited resources in term of \ac{BH} and \ac{FH}.
% In the rest of the paper, with access Link we refer to the transmission link between users and the communication technology on the \ac{UAV}, while for \ac{BH} it is considered the link between the Core Network and the UAV.
Given limited resources in terms of access and BH/FH link, the objective of this paper is to compare these three different approaches in term of data rate, latency and energy efficiency.

\subsection{Comparison of the proposed architectures}
The deployment of the proposed architectures over the sea faces challenges of complexity, energy consumption and sustainable \ac{BH} load. On one hand, the \ac{UAV} has a limited volume, weight and available energy. On the other hand, such a critical scenario has challenging demands, such as high reliability and low energy consumption.
Thus, the identification of the candidate system architecture is not trivial since it should consider the following key factors. 
Firstly, the off-shore scenario taken into consideration is critical from a communication point of view, since requires a long range \ac{BH}/\ac{FH} link %with high capacity and reliability 
to connect the \ac{UAV} to the core network. Secondly, to be efficient and use new immersive technologies, search and rescue operations demand for reliable, low latency and medium-high data rate services, i.e. access side of the network. Thirdly, the existing communication technology, such as satellite, are not be able to support primary performance targets like throughput and packet delay. 
%Finally, to support the search and rescue operations for their entire length, the \ac{UAV}'s operational time has to be prolonged saving some energy from the payload. 

%\begin{itemize}
%    \item Search and rescue operations usually take place off-shore, about $20 km $ from the coast.
%    \item Search and rescue operations demand for reliable, low latency and medium-high data rate services.% \cite{LTEPublicSafety}.
    % \item The height of the UAV would facilitate a \ac{LOS} communication with the core network and the \acp{UE}.
%    \item A long range BH/FH network with high capacity and reliability must connect the \ac{UAV} with the core network.
 %   \item The existing communication technology, such as satellite, might not be able to support primary performance targets such as throughput and packet delay.
%\end{itemize}
Now we will have a closer look at the candidate architectures:

\subsubsection{Relay \ac{UAV}} Compared to a \ac{UAV}-\ac{BS}, relays require less processing power, because their radio equipment are relatively simple. %\cite{Gesbert_UAVRelayLTE}. 
Hence, they can work with a light payload \ac{UAV}, reducing costs and power consumption. However, this solution is expected to offer less than twice the throughput a flying \ac{BS} can offer, due to the sharing of resources between communication with the \acp{UE} and relaying to the \ac{BS} \cite{Prague_Uni_Gesbert_CanUAVsubstituteSCBS}.

\subsubsection{\ac{UAV}-\acp{BS}} Flying \ac{UAV}-\acp{BS} can potentially provide enough data rate to satisfy the estimated data capacity required in emergency scenarios over the sea. %\cite{LTEPublicSafety}. 
This comes at the expense of high deployment hardware cost and weight, and a consequent reduction in the endurance of the \ac{UAV}, that might affect the search and rescue mission time.
Moreover, providing a long range \ac{BH} represents a main challenge of this configuration.

\subsubsection{\ac{RRH}-\ac{UAV}} The \ac{RRH}-\ac{UAV} approach has the advantages of a \ac{C-RAN} network. In its conventional configuration, the radio element of the base station (\ac{RRH} or \ac{RAU}) is separated from the element processing the baseband signal (\ac{BBU}), which is centralized in a single location or virtualized in the cloud.
Thus, an onboard \ac{BBU} is not required, alleviating the weight and the payload complexity on the \ac{UAV}. Moreover, the \ac{C-RAN} configuration enhances resource management and reduce power consumption on the \ac{UAV} by centralising the baseband processing.
Howewer, such a functional split requires to connect the \ac{BBU} and the flying \ac{RRH} through a long range wireless, high speed, low latency link. Conventional \ac{FH} interfaces have large overhead and may fall short in support of this architecture. Moreover, the required capacity on the \ac{FH} increase linearly if an antenna array is deployed on the \ac{UAV} to perform beamforming techniques. \ac{FH} constraints in conventional \ac{C-RAN} networks are the focus of several research works \cite{Alimi_CRANSurvey}, but the effects of these limits on \ac{UAV} communication has not been properly investigated. In fact, caching techniques have been proposed to offload the limited \ac{UAV}-\ac{BBU} \ac{FH} \cite{Saad_CachingSKy}, while other works consider unpractical fiber optical cable links or general high speed and distortion free \ac{FH} links \cite{Chiaraviglio_UAV_RuralAreas}. %\cite{Zhang_CoMPRRH}
In this paper, using a conventional FH dimensioning, we highlight the impact of FH capacity and latency requirements for search and rescue operations missions.

\subsection{Wireless BH/FH analysis for long range coverage at sea}
In terrestrial networks, when the optical fiber connectivity is unavailable, the main solution is a long range wireless BH/FH. Due to cost, latency and implementation issues, popular options like microwave point to point and satellite links cannot be considered attractive for the \ac{UAV} BH/FH link over the sea.
Consequently, we have identified different wireless technologies (Table \ref{tab:BackaulLinkComparison}) to come up with a cost effective technological option for the BH/FH link between the \ac{UAV} and the core network.

The first solution is represented by reaching the \ac{UAV} by a \ac{BS} on the shore operating at sub 6GHz, giving availability of a licensed but limited spectrum.
Wider component carrier bandwidth up to 400 MHz and low spectrum cost could be reached shifting the operating frequency to higher bands, such as \ac{mmWave} band. 
The propagation properties of the radio signal at these frequencies, together with the specific atmospheric conditions over the sea, open the way to several research challenges and opportunities:
the high level of humidity leads to an additional free space absorption \cite{Rappaport_mmWaveOverview} but, at the same time, atmospheric ducts, result of different refractive indices of the layers over the water, can be exploited to propagate the signal over longer distance \cite{UAV_mmWAve_Ducting_Zhang}.
Although mentioned, the \ac{mmWave} option is not focus of this paper, and it is left for future investigation.
% at this band the attenuation at free space is low and the coverage range is wide\cite{5G\ac{BH}_Siddique}.
% The over the sea environment specific atmospheric conditions,where humidity leads to an additional atmospherical absorption \cite{Rappaport_mmWaveOverview}
% the high free space path loss at these frequencies together with the specific conditions over the sea, where humidity leads to an additional atmospherical absorption, limits the reliability and coverage of this option.
In combination with the operating frequency, to increase the range of the over the sea \ac{BH} solutions, we have considered to employ multiple antennae on the transmitter.
In fact, the use of multiple antennae in transmission in \ac{LOS}, although it might not achieve \ac{MIMO}'s capacity-multiplying effect, still enhances the coverage range. %\cite{MIMO_LOS_Bohagen}.
As last, we mention also the \ac{FSO}. Even if promising for its high data rate and low spectrum cost, this technology is very sensitive to inclement weather conditions. %\cite{FSOVerticalFH_Alzenad}
%  and the sea rough surface  the path loss models presented in in \ref{ChannelModelAL} unable to model the channel over the sea surface \cite{Rappaport_mmWaveOverview}. 
\begin{table*}[ht]
\centering
 \caption{BH/FH Link communication technologies over the sea}\label{tab:BackaulLinkComparison}
 \setlength\tabcolsep{4pt}
 \begin{tabularx}{\textwidth}{|c|X|X|l|c|c|}
        \hline
        \emph{Technology}  & Benefits & Challenges &\emph{Data Rate} & Bandwidth  \\
        \hline
        sub6GHz SISO & Long range, Robust to over-the-sea atmospheric conditions  & Limited Spectrum & $C = B log_2 (1+SNR)$ & 20 MHz \\
        \hline
        sub6GHz - 2x2 \ac{MIMO} & Increase in coverage range & LOS Conditions are particularly stressy %\cite{marzetta_larsson_yang_ngo_2016} 
        & $C = 2B log_2 (1+SNR)$ & 20 MHz\\
        \hline
        sub6GHz - massive \ac{MIMO} &  Array gain for range extension & 
         Correlation between antenna elements when transmitting at long range & $C = B K (1-\frac{K}{\tau}) log_2 (1+\frac{c_{csi}M SNR}{K SNR + 1})$ %\cite{10mythsMMIMO} 
         & 20 MHz \\
        \hline
        Free space Optics & Potential high data rate & Greatly affected by weather conditions. Sensitive to Pointing and Vibration loss,
        Higher cost & $C = \frac{P_{TX} \eta_r \eta_t 10^{-L_{pol}/{10}} 10^{-L_{atm}/{10}} A_R}{A_B E_P N_b}$
        & / \\
        \hline
        mmWave  & High availability of spectrum, ducting effect increase the coverage range & Sensitive to multiple over-the-sea atmospheric conditions, existing path loss model over the sea are valid only for short range %\cite{NovelMaritimelChannelMOdels_Mehrnia}  
        & $C = B log_2 (1+SNR)$  & 400 MHz\\
        \hline
        % Microwave Point to Point  & High data rate & must be in LOS, heavy antenna required & / & / \\
        % \hline
        % Satellite  & Wide coverage, supports mobility & Low data rate, high cost & / & / \\
        % \hline
\end{tabularx}
\end{table*}

\section{System model and problem formulation}
Let us consider a set of users ${U = 1, 2, ...,N_U}$ in an area A over the sea that cannot be served by any terrestrial \ac{BS}.
The $N_U$ users are split and randomly concentrated in two different areas of same diameter r, representing the people in distress and the rescue team.
A \ac{UAV} is hovering at altitude $h$ to transmit signals to the users on the sea.
 We assume that the main factor which affects the service quality offered by the \ac{UAV} is pathloss, as the \ac{UAV} operates at \ac{LOS} with no \ac{DL} interference from other cells.
In order to derive the data rate and the latency for the three different system architectures, it is imperative to find an air to sea path loss model.

% %write what can make a \ac{BH} going out temporary
% The correct functioning of the \ac{UAV} can be influenced by the battery

% %how to solve it

% Consider a \ac{UAV} hovering at altitude h an area A over the sea. The UAV is considered to have only one antenna for transmission and reception its radio access operating frequency is the \ac{LTE} frequency band. Since the presence of only one UAV and the long distance from the ground base stations, we consider no interference on the downlink.
%  a set $U$ of users spread across a $L \times L$ square area. A portion of the users is assumed to be randomly concentrated in an area of diameter 10m, representing the people in distress, and another portion of users is assumed to be randomly concentrated in another area of 10m diameter, representing the rescue team. Thus, two type of users are considered: either users that belong to an emergency team or regular users. For both type, users are considered to have low or no mobility.
 
\subsection{Air to Sea Channel Model}\label{ChannelModelAL}
Modeling wireless channels for over the sea wave propagation is a relatively new research topic. 
The UAV-to-ground path loss is generally computed using the urban model in \cite{LAPpathLoss}, that adds to the free space loss an additional one which depends on whether there is \ac{LOS} between the \ac{UAV} and the users or not.
Since the unique characteristics of over the water environment, from the roughness of the sea surface to the specific atmospheric phenomena occurring over the sea (humidity, evaporation ducting), this model might be inadequate. % such as the high humidity level and the evaporation ducting phenomena,
A robust wireless channel model for over the sea wave propagation has not been developed yet and the existing approaches vary with frequencies, technology used and implementation scenario.
% Th path loss of the UAV-to-ground communication link is generally computed using the urban model $PL_d = 20 log_{10}(\frac{4 pi f_c d}{c}) + \epsilon$ \cite{LAPpathLoss}, where d is the distance between drones and users, c is the speed of the light. This model, that to the free space loss adds an additional loss ($\epsilon$) which depends on whether there is \ac{LOS} between the \ac{UAV} and the users or not, might be inadequate for this specific over the water environment.
% As first, for air-to-ground wireless links over the sea, a \ac{NLOS} link is very unlikely. %Thus, for open water settings, the path loss generally follows the free space curve \cite{Matolak_OverWater_AGChannelModel}.

In this paper, as in the most of channel models in literature,  we consider the sea surface to be smooth and we do not consider any ducting effect. Consequently, we adopt a two ray model to take into consideration the reflection component of the sea. % and fits better than a free space or log distance, particularly as distance increases \cite{Matolak_OverWater_AGChannelModel}.
% Consequently, we take into account the reflection component of the sea, that is more prominent at lower frequency since the sea is smoother at longer wavelength. For this reason, the two-ray model, particularly the Curved earth ray version fits better than a free space or log-distance or flat-earth two ray model, particularly as distance increases \cite{Matolak_OverWater_AGChannelModel}. that is visible at large values of distance ($\geq 10km $)
The path loss can be expressed as follows:
\begin{equation}
    L(d) = -20 log_{10}\{\frac{\lambda}{4 \pi d} [2sin(\frac{2 \pi h_t h_r }{\lambda d})] \},
\end{equation}
where d is the distance between the \ac{UAV} and the user.

\subsection{Transmission Model}
To compare the \ac{DL} channel capacity at user $i\, \epsilon \ U = \{1, ..., N_U\}$ served by the \ac{UAV} network for each of the proposed network architectures, we have computed the channel capacity on both ground \ac{BS}/\ac{BBU}-UAV, and \ac{UAV}-users transmission links.
\subsubsection{UAV-User Links} given the channel model presented in \ref{ChannelModelAL}, the average path loss depends on the location and the height of the user i and the UAV. % considered in \ac{LOS}. 
Based on the path loss, for a generic user i, it is possible to compute the power received, $P_R$. 
Recalling that the average \ac{SNR} at user i can be expressed as $\ac{SNR} =P_R / N$, where N is the noise power, the corresponding \ac{DL} access data rate for each network architecture is shown in Table \ref{tab:AccessLinkComparison}.
% Since the available bandwidth on the radio access is equally divided between the $N_U$ users served by the UAV, the potential total access capacity is $C_{Access} = \sum_{i}^{N_U} C_{i_{Access}}$ where $C_{i_{Access}}$ is the \ac{DL} capacity of the single user.
\subsubsection{Ground \ac{BS}/\ac{BBU}-UAV Links}
For the \ac{BH} and \ac{FH} link to the UAV, we consider a fixed distance $d$ much larger than the distance \ac{UAV}-users, to represent the offshore situation of the \ac{UAV}.
The corresponding \ac{DL} backhauling/fronthauling rate at the \ac{UAV} depends in this case on the \ac{LOS} path loss and the technology deployed on the shore \ac{BS}/\ac{BBU} (Table \ref{tab:BackaulLinkComparison}).
%In our transmission model we assume that the backhaul capacity is equally divided between the $N_U$ users served by the UAV.
%The total downlink backhaul rate provide to the \ac{UAV}, $C_{Backhaul}$, cannot be smaller than the actual agrregated downlink access rate provided by the

% Given this channel model, for the system architectures I, II and III, it is possible to compute the channel capacity on the backhaul, $C_{Backhaul}$, and on the access link for each user $i\, \epsilon \ U = \{1, ..., N_U\}$ served by the \ac{UAV} network, $C_{i_{Access}}$. Since the available bandwidth on the radio access is equally divided between the $N_U$ users served by the UAV, the potential total access capacity is $C_{Access} = \sum_{i}^{N_U} C_{i_{Access}}$ where $C_{i_{Access}}$ is the DL capacity of the single user. The actual aggregated downlink access rate provided by the \ac{UAV} cannot be larger than the available backhaul rate, which leads to $\{C_{Access} \leq C_{Backhaul}\}.$ %\cite{bonfante20185g}, \cite{EURECOMBackhaul_TechReport}. % $\{C_{Access} \leq C_{Fronthaul} \}$
% In our transmission model we assume that the backhaul capacity is equally divided between the $N_U$ users served by the UAV.

\subsection{Latency Model}
The latency experienced by the user $i$ served by the \ac{UAV} wireless network is the sum of three components:
the latency on the \ac{BH} or \ac{FH} link, the latency on  the access link and the processing latency (or computational) due to the time consumed to process the radio signal on the \ac{UAV}.
The resulting combination of the factors mentioned above for the three system architectures considered in this paper are summarized in Table \ref{tab:AccessLinkComparison}. %Inspired by the work in \cite{Saad_Beyond5G} and 
% Following the three proposed architectures, we have computed the latency as follows:
% \begin{enumerate}[label=\Roman*]
%     \item Latency Relay Link 1 - Latency Relay Link 2 - Relay Computation Latency
%     \item Latency Backhaul link - Latency AccessLink - BS Computation Latency
%     \item Latency FH link - Latency AccessLink  - RRH Computation Latency
% \end{enumerate}
The link connecting the cloud with the core network is assumed to be an ideal high data-rate fiber link and, therefore, its latency is considered negligible in respect of the total computation.
% Latency AccessLink is the latency due to the transmission of the data in the access link and is the same for both the \ac{RRH} and\ac{UAV} approach, while for the BH/FH link the latency depends on the actual achievable throughput and differs from one configuration and the other.
The \ac{RRH} computation latency includes only the processing delay given by the RF part, while for the \ac{BS} all the parts (e.g. demodulation, decoding/coding etc) are involved in the processing \cite{AHmadiBBU_RRH_distance}.
% % Thus, the latency at the user i can be  determined as follows:

% % \begin{equation}
% %  \tau_i =
% % \begin{cases}
% %      \frac{L}{C_{FH}} + \frac{L}{C_{i_{Access}}} + tau_{comp_{RRH}}, &\text{for Link1}\\
% %      \frac{L}{C_{FH}} + \frac{L}{C_{i_{Access}}} + tau_{comp_{BS}}, &\text{for Link2}\\
% %      \frac{L}{C_{FH}} + \frac{L}{C_{i_{Access}}}, &\text{for Link3}
% % \end{cases}    
% % \end{equation}

\subsection{Energy efficiency model}
%According to the most recent model for energy consumption in the \ac{UAV} communication\cite{SurveyGalati_UAV}, 
The available energy on the \ac{UAV} is consumed by both communications (electronics) and mobility (mechanical) but the ratio of communication energy consumption to the total energy consumption of \acp{UAV} is generally negligible \cite{SurveyGalati_UAV}. 
As a result, \ac{UAV}'s altitude and payload weight play the most important role in the total power consumption. 
Considering to use the same \ac{UAV} for all the three architectures, %but RRH and relay would enable to use lighter
in Table \ref{tab:AccessLinkComparison} we include an indicative weight related to the communication technology deployed on the \ac{UAV}. 
%this weight influence the choice of the UAV (usually small UAV \cite{SurveyGalati_UAV} are fro BS) and the total architecture weight and can be considered a comparison factor

\begin{table*}[t]
 \centering
 \caption{Communication Technologies Comparison}\label{tab:AccessLinkComparison}
 \resizebox{0.9\textwidth }{!}{ 
    \begin{tabular}{llll}
        \hline
        \emph{Aerial Platform} & \emph{Data Rate} & \emph{Latency} &  \emph{Weight}\\ %\emph{Energy model} & 
        \hline
        Relay-UAV & $C = (B/2) log_2 (1+SNR)$ & \small{Relay Link 1 + Relay Link 2 +  Relay ComputationLatency}  &  400g  \\ %$E = (\beta + \alpha h)*t + P_{max}(h/v)$ & 
        BS-UAV & $C = B log_2 (1+SNR)$ & \small{AccessLink + BackhaulLink + BS ComputationLatency} &  $\geq 10 kg$ \\ %&$E = (\beta + \alpha h)*t + P_{max}(h/v)$  
        RRH-UAV &  $C = B log_2 (1+SNR)$  & \small{AccessLink + FronthaulLink + ComputationLatency}    & $\leq 6 kg $ \\ %& $E = (\beta + \alpha h)*t + P_{max}(h/v)$
        \hline
    \end{tabular}}
\end{table*}

\subsection{A suboptimal BH/FH-UAV technology combination}
Defining a cost function is a common approach to deal with the choice of a suboptimal solution. The set of \ac{BH} technologies options $i = \{1, 2, 3, 4\} = \{$sub6GHz SISO, sub6GHz - 2x2 \ac{MIMO}, sub6GHz - massive \ac{MIMO}, Free Space Optics$\}$ and the proposed \ac{UAV} architecture options $j = \{1,2,3\} = \{$BS, RRH, Relay$\}$ lead to a finite number of combinations $ij$.
First, the most important indicators for an emergency \ac{UAV} wireless network over the sea can be defined, such as data rate, latency and energy efficiency. If each of the identified attribute is normalized and weighted, the cost function can give an indication of the efficiency of the $ij$ combination to fulfill the requirements of the search and rescue operation over the sea. In this paper normalized values are defined in a way which the lower value is more desirable. The cost function can be expressed then as
\begin{equation}
  CF_{ij} = \sum_{z} w_z s_z,\;\;\; z= \{1,..,3\}  
\end{equation}
where $s_z$ is the attribute considered and $w_z$ is the corresponding weight coefficient.
 % As an example, if we assume $w_1 = 1$ and $w_2, w_3 = 0 $, corresponding to the case where priority is given to the data rate, the utility utility score of the cost function leads to a normalized score of 0.618421053 for the \ac{BS}-\ac{UAV} backhauled by a 2x2 MIMO \ac{BS} on the shore, that represent the best choice to adopt in case of a high data rate requirement.

\section{Simulation and numerical results}
In order to show effectively the difference for the proposed architectures, a simulation scenario has been built in MATLAB.
A summary of the simulation parameters is shown in Table \ref{tab:SimParams}.
\begin{table}[ht]
 \centering
 \caption{\footnotesize{Simulation Parameters}}\label{tab:SimParams}
    \footnotesize 
    \begin{tabular}{|p{0.5\columnwidth}|p{0.3\columnwidth}|}
        \hline
        \emph{Parameters} & \emph{Value}\\
        \hline
        Side of the square area A & 1 Km\\
        Side of the users area & 100m\\
        Number of users spots & 2\\
        Number of UAVs & 1\\
        UAV height & 200m\\
        Ratio of rescuers over the users &  1/3 $\div$ 2/3\\
        User height & 2m\\
      % Ratio of rescue team users & Passive\\
        Ground BS EIRP & 43 dBm\\
        UAV EIRP & 20 dBm\\
        Access Link Bandwidth &  20 MHz\\
        Access Link carrier frequency & 2.6 GHz\\
        % High SNR requirement & \\
        % Low SNR requirement & \\
        Number of independent runs & 100\\
        \hline
    \end{tabular}
\end{table}

\begin{figure}[ht]
    \centering
    \includegraphics[width=1\columnwidth]{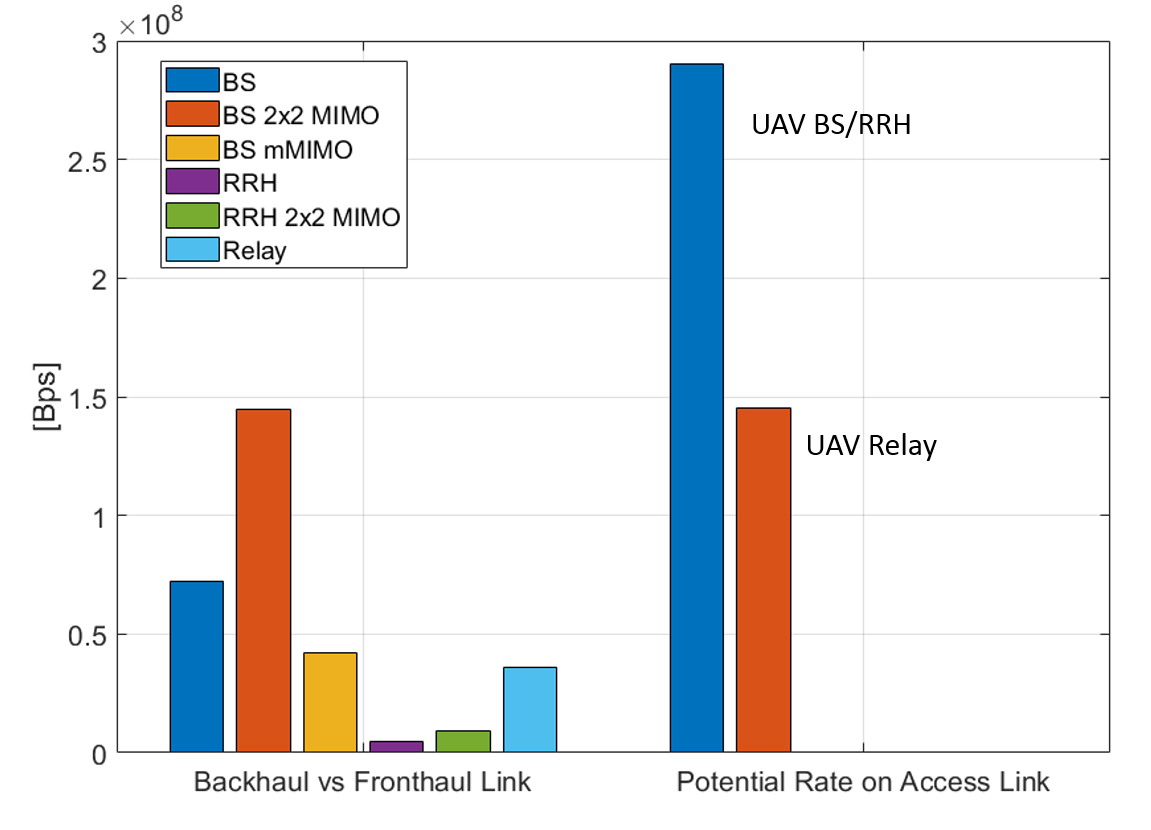}
    \caption{Data rate available at the UE based on different \ac{BH}/\ac{FH} solutions: the potential aggregated \ac{DL} access rate provided by the \ac{UAV} is limited by the available \ac{BH}/\ac{FH} rate.}
    \label{fig:Data_rate_results}
\end{figure}

\begin{figure}[ht]
    \centering
    \includegraphics[width=1\columnwidth]{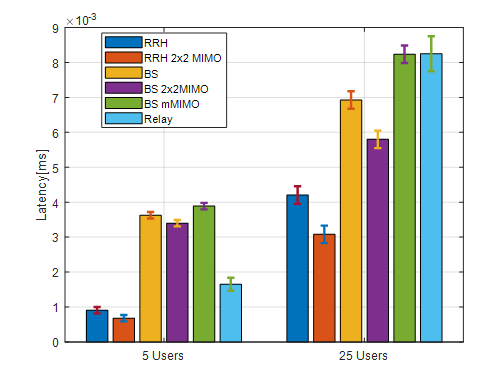}
    \caption{Average total latency for the proposed system architectures with increasing number of UEs on the access link }
    \label{fig:latency}
\end{figure}

Different BH/FH options have been considered to provide long range BH/FH to the \ac{UAV}.
The left side of Figure \ref{fig:Data_rate_results} shows the average \ac{DL} BH/FH channel capacity for different technologies.
The actual total \ac{DL} access rate deliverable to the \acp{UE} by the \ac{UAV} cannot be larger than the BH/FH \ac{DL} rate. Thus, as visible comparing the BH/FH rate with the only potential data rate on the access link (on the right in figure \ref{fig:Data_rate_results}), the distance from the ground makes hard to provide the \ac{UAV} a BH/FH link without limiting the performance of the system.
Figure \ref{fig:latency} shows the simulation latency results for different numbers of users. As expected, due to the lower computation time on the UAV, the \ac{RRH} and the Relay configuration give significant benefits in term of latency in respect to the \ac{BS} configuration.
These results emphasizes the fact that implementing a \ac{UAV}-aided wireless network over the sea struggle with issues at different levels. 
Different evaluation parameters lead to opposite results.
For this reason an optimal solution is hard to design, but, considering different performance indicators it is possible to choose the best combination of \ac{BH} technology and system architecture.

\section{Conclusion}
In this paper we focused on the problem of providing cellular connectivity to offshore areas for search and rescue operations. We have presented three different system architectures and highlighted the main benefits and challenges. 
To evaluate the proposed architectures, we have compared their performance in term of data rate, latency and energy efficiency. Results shows the need of a cost function to evaluate the best architecture based on an assigned priority.

\section*{Acknowledgment}
{\small This material is based upon works supported by the Irish Research Council under Grant GOIPG/2017/1741.}

\begin{acronym} 
\acro{5G}{Fifth Generation}
\acro{ACO}{Ant Colony Optimization}
\acro{BB}{Base Band}
\acro{BBU}{Base Band Unit}
\acro{BER}{Bit Error Rate}
\acro{BH}{Backhaul}
\acro{BS}{Base Station}
\acro{BW}{bandwidth}
\acro{C-RAN}{Cloud Radio Access Networks}
\acro{CAPEX}{Capital Expenditure}
\acro{CoMP}{Coordinated Multipoint}
\acro{CPRI}{Common Public Radio Interface }
\acro{DAC}{Digital-to-Analog Converter}
\acro{DAS}{Distributed Antenna Systems}
\acro{DBA}{Dynamic Bandwidth Allocation}
\acro{DL}{Downlink}
\acro{FBMC}{Filterbank Multicarrier}
\acro{FEC}{Forward Error Correction}
\acro{FH}{Fronthaul}
\acro{FFR}{Fractional Frequency Reuse}
\acro{FSO}{Free Space Optics}
% \acro{GA}{Genetic Algorithms}
\acro{GS}{Ground Station}
\acro{GSM}{Global System for Mobile Communications}
\acro{HAP}{High Altitude Platform}
\acro{HL}{Higher Layer}
\acro{HARQ}{Hybrid-Automatic Repeat Request}
\acro{KPI}{Key Performance Indicator}
\acro{IoT}{Internet of Things}
\acro{LAN}{Local Area Network}
\acro{LAP}{Low Altitude Platform}
\acro{LL}{Lower Layer}
\acro{LOS}{Line of Sight}
\acro{LTE}{Long Term Evolution}
\acro{LTE-A}{Long Term Evolution Advanced}
\acro{MAC}{Medium Access Control}
\acro{MAP}{Medium Altitude Platform}
\acro{ML}{Medium Layer}
\acro{MME}{Mobility Management Entity}
\acro{mmWave}{millimeter Wave}
\acro{MIMO}{Multiple Input Multiple Output}
\acro{NFP}{Network Flying Platform}
\acro{NFPs}{Network Flying Platforms}
\acro{NLOS}{Non Line of Sight}
\acro{OFDM}{Orthogonal Frequency Division Multiplexing}
\acro{PAM}{Pulse Amplitude Modulation}
\acro{PAPR}{Peak-to-Average Power Ratio}
\acro{PGW}{Packet Gateway}
\acro{PHY}{physical layer}
\acro{PSO}{Particle Swarm Optimization}
\acro{PTP}{Poin to Point}
\acro{QAM}{Quadrature Amplitude Modulation}
\acro{QoE}{Quality of Experience}
\acro{QoS}{Quality of Service}
\acro{QPSK}{Quadrature Phase Shift Keying}
\acro{RF}{Radio Frequency}
\acro{RN}{Remote Node}
\acro{RAU}{Remote Access Unit}
\acro{RAN}{Radio Access Network}
\acro{RRH}{Remote Radio Head}
\acro{RRC}{Radio Resource Control}
\acro{RRU}{Remote Radio Unit}
\acro{RSRP}{Reference Signals Received Power }
\acro{SCBS}{Small Cell Base Station}
\acro{SDN}{Software Defined Network}
\acro{SNR}{Signal-to-Noise Ratio}
\acro{SON}{Self-organising Network}
\acro{TETRA}{Trans-European Trunked Radio}
\acro{TDD}{Time Division Duplex}
\acro{TD-LTE}{Time Division LTE}
\acro{TDM}{Time Division Multiplexing}
\acro{TDMA}{Time Division Multiple Access}
\acro{UE}{User Equipment}
\acro{UAV}{Unmanned Aerial Vehicle}

\end{acronym}

% Usage in text:
% \ac{BBU} for singular
% \acp{BBU} for plural

\bibliographystyle{IEEEtran}
{\footnotesize  
\bibliography{References.bib}
}
\end{document}